\documentstyle[12pt,epsfig,graphics,cite,amssymb,amsmath,bm,booktabs,dcolumn]{article}
\newcolumntype{d}[1]{D{.}{.}{#1}}
\begin{document}\bibliographystyle{plain}\begin{titlepage}
\renewcommand{\thefootnote}{\fnsymbol{footnote}}\hfill
\begin{tabular}{l}HEPHY-PUB 999/17\\December 2017\end{tabular}\\[2cm]
\Large\begin{center}{\bf GOLDSTONIC PSEUDOSCALAR MESONS\\IN
BETHE--SALPETER-INSPIRED SETTING}\\[1cm]\large{\bf Wolfgang
LUCHA\footnote[1]{\normalsize\ {\em E-mail address\/}:
Wolfgang.Lucha@oeaw.ac.at}}\\[.3cm]\normalsize Institute for High
Energy Physics,\\Austrian Academy of Sciences,\\Nikolsdorfergasse
18, A-1050 Vienna, Austria\\[1cm]\large{\bf Franz
F.~SCH\"OBERL\footnote[2]{\normalsize\ {\em E-mail address\/}:
franz.schoeberl@univie.ac.at}}\\[.3cm]\normalsize Faculty of
Physics, University of Vienna,\\ Boltzmanngasse 5, A-1090 Vienna,
Austria\\[2cm]{\normalsize\bf Abstract}\end{center}\normalsize
\noindent For a two-particle bound-state equation closer to its
Bethe--Salpeter origins than Salpeter's equation, with
effective-interaction kernel deliberately forged such as to
ensure, in the limit of zero mass of the bound-state constituents,
the vanishing of the arising bound-state~mass, we scrutinize the
emerging features of the lightest pseudoscalar mesons for their
agreement with the behaviour predicted by a generalization of the
Gell-Mann--Oakes--Renner relation.\vspace{3ex}

\renewcommand{\thefootnote}{\arabic{footnote}}\end{titlepage}

\section{Stimulus: Janiform Nature of Pseudoscalar Mesons}Within
the spectrum of known elementary particles, the lowest-mass
pseudoscalar mesons, the pions and kaons, occupy a unique
position: On the one hand, they are understood to be
quark--antiquark bound states. On the other hand, they should form
the pseudo-Goldstone bosons the presence of which is necessitated
by the spontaneous (and, in addition, explicit) breaking of the
chiral symmetries of quantum chromodynamics (QCD). Their
ambivalence renders their quantum-field-theoretic description by
the Bethe--Salpeter formalism \cite{BS,GML,SB}~or by suitable
three-dimensional reductions of the latter, such as Salpeter's
equation \cite{SE} or the less serious simplification embodied by
the bound-state equation of Ref.~\cite{WL05:LS}, a delicate~task.

Recently, by means of a dedicated inversion technique \cite{WL13},
we embarked on a systematic exploration of the effective
interactions between bound-state constituents entering, in form of
configuration-space central potentials $V(r),$ $r\equiv|\bm{x}|,$
in Salpeter's approach \cite{WL13p,WL15,WL16:ARP,WL16:DSE} or our
softened approximation \cite{WL16i,WL16@,WL16c}. Here, we
demonstrate how the reliability of predictions for such a meson is
judged by their fulfilment of Gell-Mann--Oakes--Renner-type
relationships.

There is no need to announce in detail the expectable outline of
this paper: By applying the ``Bethe--Salpeter-inspired''
bound-state formalism constructed in Ref.~\cite{WL05:LS}
(Sec.~\ref{Sec:IL}) to the lowest pseudoscalar mesons
(Sec.~\ref{Sec:SS}), some of their thus predicted basic features
(Sec.~\ref{Sec:MP}) are examined, for the Goldstone-friendly
interquark potential derived in Ref.~\cite{WL16i}
(Sec.~\ref{Sec:I}), with regard to their compatibility with
relationships of Gell-Mann--Oakes--Renner type
(Sec.~\ref{Sec:GOR}). Throughout this paper, we use the natural
units of relativistic quantum physics:~$\hbar=c=1.$

\section{Bethe--Salpeter Formalism: an Instantaneous Limit}
\label{Sec:IL}In principle, the Bethe--Salpeter formalism
\cite{BS,GML,SB} is the proper tool for a
\emph{Poincar\'e-covariant\/} description of bound states within
the framework of relativistic quantum field theories: the
homogeneous Bethe--Salpeter equation relates the Bethe--Salpeter
amplitude, encoding the distribution of the relative momenta of
the bound-state constituents, to, for $n$ constituents, their $n$
propagators and the ($2\,n$)-point Green function encompassing all
their interactions.

We intend to deal with mesons, so we let $n=2$ and focus to bound
states composed of a fermion and an antifermion. In terms of the
individual coordinates and momenta of the two bound-state
constituents, discriminated by the subscript $i=1,2$, parameters
$\eta_{1,2}\in\mathbb{R},$~and the center-of-momentum and relative
coordinates and total and relative~momenta given~by
\begin{xalignat*}{3}X&\equiv\eta_1\,x_1+\eta_2\,x_2\ ,&x&\equiv
x_1-x_2\ ,&\eta_1+\eta_2&=1\ ,\\P&\equiv p_1+p_2\ ,&p&\equiv
\eta_2\,p_1-\eta_1\,p_2\ ,&P^2&=\widehat M^2\ ,\end{xalignat*}the
momentum-space Bethe--Salpeter amplitude of each such bound state
$|{\rm B}(P)\rangle$ is defined by the Fourier transform of the
vacuum-to-bound-state matrix element of the time-ordered product
of the field operators $\psi_i(x_i)$ of both involved bound-state
constituents according~to$$\Phi(p,P)\equiv\exp({\rm i}\,P\,X)
\int{\rm d}^4x\exp({\rm i}\,p\,x)\,\langle 0|{\rm T}(\psi_1(x_1)
\,\bar\psi_2(x_2))|{\rm B}(P)\rangle\ .$$By Poincar\'e covariance,
the propagator of any fermion $i$ is fixed by only two
Lorentz-scalar functions, interpretable as, {\it e.g.}, its mass
$M_i(p^2)$ and wave-function renormalization~$Z_i(p^2)$:
\begin{equation}S_i(p)=\frac{{\rm i}\,Z_i(p^2)}{\not\!p-M_i(p^2)
+{\rm i}\,\varepsilon}\ ,\qquad\not\!p\equiv p^\mu\,\gamma_\mu\
,\qquad\varepsilon\downarrow0\ ,\qquad i=1,2\ .\label{Eq:P}
\end{equation}

As a relativistic formalism, however, the Bethe--Salpeter approach
to bound states is, in general, threatened by the (possible)
appearance of excitations in the relative-time variable of the
bound-state constituents. Lacking nonrelativistic counterparts,
the interpretation of such solutions poses a severe challenge.
This obstacle to the light-hearted application of the
Bethe--Salpeter approach to relativistic bound-state problems
triggered the construction of approximations not plagued by
timelike excitations: By seeking suitable three-dimensional
reductions of the Poincar\'e-covariant framework by constructing
instantaneous limits of the homogeneous Bethe--Salpeter formalism,
bound-state equations for the Salpeter amplitude$$\phi(\bm{p})
\equiv\int\frac{{\rm d}p_0}{2\pi}\,\Phi(p,P)$$have been proposed.
The supposedly simplest among all these is the one due to Salpeter
\cite{SE}, who assumed, for all bound-state constituents, both
constant masses and free propagation.

Some time ago, in an attempt to overcome some of the limitations
induced by Salpeter's assumption \cite{SE} of strict constancy of
the masses of all bound-state constituents by retaining as much as
feasible from the dynamical information encrypted in the momentum
behaviour of (fermion) propagator functions, we formulated an
approximation \cite{WL05:LS} to the homogeneous Bethe--Salpeter
equation which is characterized by the ignorance of merely the
dependence of both propagator functions on the time components of
the involved momentum variables. Recalling free Hamiltonian, free
energy and energy projection operators of fermion~$i=1,2,$
$$H_i(\bm{p})\equiv\gamma_0\,[\bm{\gamma}\cdot\bm{p}+M_i(\bm{p}^2)]\
,\qquad E_i(\bm{p})\equiv\sqrt{\bm{p}^2+M_i^2(\bm{p}^2)}\
,\qquad\Lambda_i^\pm(\bm{p})\equiv\frac{E_i(\bm{p})\pm
H_i(\bm{p})}{2\,E_i(\bm{p})}\ ,$$the resulting instantaneous-limit
equation \cite{WL05:LS} governing fermion--antifermion bound
states, with three-momentum $\bm{P}$ and mass generically labelled
by $\widehat M,$ can be cast \cite{WL07H,WL16i,WL16c,WL10@}, in
the center-of-momentum frame of such bound states, that is, for
$\bm{P}=\bm{p}_1+\bm{p}_2=0$,~into~the~form
\begin{equation}\phi(\bm{p})=Z_1(\bm{p}^2)\,Z_2(\bm{p}^2)\left(
\frac{\Lambda_1^+(\bm{p})\,\gamma_0\,I(\bm{p})\,
\Lambda_2^-(\bm{p})\,\gamma_0}{\widehat M-E_1(\bm{p})-E_2(\bm{p})}
-\frac{\Lambda_1^-(\bm{p})\,\gamma_0\,I(\bm{p})\,
\Lambda_2^+(\bm{p})\,\gamma_0}{\widehat M+E_1(\bm{p})+E_2(\bm{p})}
\right),\label{Eq:CMF}\end{equation}where an integral kernel
$K(\bm{p},\bm{q})$ subsumes the, by assumption instantaneous,
interactions:\begin{equation}I(\bm{p})\equiv\int\frac{{\rm
d}^3q}{(2\pi)^3}\, K(\bm{p},\bm{q})\,\phi(\bm{q})\
.\label{Eq:I}\end{equation}

The normalization of any state $|B(\bm{P})\rangle$ may be defined
by a normalization factor $N(P);$ popular choices of $N(P)$ are
relativistically noncovariant normalizations such as $N(P)=1$ and
$N(P)=(2\pi)^3,$ or a relativistically covariant normalization
such as $N(P)=(2\pi)^3\,2\,P_0$:\begin{equation}\langle
B(\bm{P})|B(\bm{P}')\rangle=N(P)\,\delta^{(3)}(\bm{P}-\bm{P}')\
.\label{Eq:BN}\end{equation}Bound-state normalization $N(P)$ and
Salpeter-amplitude norm $\|\phi\|$ are related via
\cite{LY+,L92I,RMMP94,OVW95}$$\|\phi\|^2=\frac{N(P)}{(2\pi)^3}\
.$$Now, neglecting in the covariant normalization condition for
the Bethe--Salpeter amplitude $\Phi(p,P)$ the dependence of the
Bethe--Salpeter interaction kernel on the bound state's total
four-momentum $P$ or assuming that, in the center-of-momentum
frame of the bound~state,\pagebreak\ the instantaneous version of
the interaction kernel does not depend on $P,$\footnote{Needless
to say, such concept of instantaneity of some interaction is
hardly compatible with relativistic covariance \cite{CHLS}.
Notwithstanding this, for the sake of peace and comparability with
other investigations,~let us bite the bullet and ignore any
potential contributions of the interaction kernel to the Salpeter
norm~$\|\phi\|.$} the norm of each Salpeter amplitude
$\phi(\bm{p})$ is given by ({\it e.g.},
Refs.~\cite[Eq.~(2.9)]{L92I}, \cite[Eq.~(29)]{RMMP94} or
\cite[Eq.~(9)]{OVW95})\begin{align*}
\|\phi\|^2&=\frac{1}{2}\int\frac{{\rm d}^3p}{(2\pi)^3}\,{\rm Tr}\!
\left[\phi^\dag(\bm{p})\left(\frac{H_1(\bm{p})}{E_1(\bm{p})}\,\phi(\bm{p})
-\phi(\bm{p})\,\frac{H_2(-\bm{p})}{E_2(\bm{p})}\right)\right]\\
&=\int\frac{{\rm d}^3p}{(2\pi)^3}\,{\rm Tr}\!
\left[\phi^\dag(\bm{p})\,\frac{H_1(\bm{p})}{E_1(\bm{p})}\,\phi(\bm{p})
\right],\end{align*}where the trace clearly extends over all
colour, flavour, and spinor degrees of freedom of the bound-state
constituents. In view of the fact that our mesonic bound states,
$|B\rangle,$ are colour singlets, this normalization of the
Bethe--Salpeter amplitude $\Phi(p,P)$ with respect to colour
induces, in the general case of $N_{\rm c}$ colour degrees of
freedom, an overall colour~factor~$1/\sqrt{N_{\rm c}}.$

We assume a kind of flavour symmetry of all fermion propagators in
our Bethe--Salpeter formalism, manifesting in form of equality of
the Lorentz-scalar functions of the same kind:
$$Z_1(\bm{p}^2)=Z_2(\bm{p}^2)=Z(\bm{p}^2)\ ,\qquad
M_1(\bm{p}^2)=M_2(\bm{p}^2)=M(\bm{p}^2)\ .$$This, in turn, implies
for the Dirac Hamiltonians, kinetic energies and projection
operators\begin{align*}H_1(\bm{p})=H_2(\bm{p})=H(\bm{p})&\equiv
\gamma_0\,[\bm{\gamma}\cdot\bm{p}+M(\bm{p}^2)]\ ,\\[1ex]
E_1(\bm{p})=E_2(\bm{p})=E(\bm{p})&\equiv\sqrt{\bm{p}^2+M^2(\bm{p}^2)}\
,\\[1ex]\Lambda_1^\pm(\bm{p})=\Lambda_2^\pm(\bm{p})=\Lambda^\pm(\bm{p})
&\equiv\frac{E(\bm{p})\pm H(\bm{p})}{2\,E(\bm{p})}\ ,\end{align*}
and serves to simplify our fermion--antifermion bound-state
equation (\ref{Eq:CMF}) a little bit further:\begin{equation}
\phi(\bm{p})=Z^2(\bm{p}^2)\left(\frac{\Lambda^+(\bm{p})\,\gamma_0
\,I(\bm{p})\,\Lambda^-(\bm{p})\,\gamma_0}{\widehat M-2\,E(\bm{p})}
-\frac{\Lambda^-(\bm{p})\,\gamma_0\,I(\bm{p})\,\Lambda^+(\bm{p})\,
\gamma_0}{\widehat M+2\,E(\bm{p})}\right).\label{Eq:F}
\end{equation}The latter variant of instantaneous Bethe--Salpeter
equation shall be used, in the following, to study those basic
features of the pseudo-Goldstone-type pseudoscalar mesons that
enter in the Gell-Mann--Oakes--Renner relation \cite{GMOR} or a
more modern generalization thereof \cite{MRT}.

\section{Spin-Singlet Fermion--Antifermion Bound Systems}
\label{Sec:SS}By definition, any \emph{pseudoscalar\/} bound state
of a spin-$\frac{1}{2}$ fermion and a spin-$\frac{1}{2}$
antifermion~is characterized by zero bound-state spin $J,$ $J=0,$
and negative parity $P,$ $P=-1.$ Thus, any such state perforce has
zero relative orbital angular momentum $\ell,$ $\ell=0,$ and
zero~total spin $S,$ $S=0$ ({\it e.g.}, Ref.~\cite{WL91}). The
latter, in turn, implies that any such state must have~positive
charge-conjugation parity $C,$ $C=+1,$ while the product of $P$
and $C$ is~negative: $C\,P=-1;$ necessarily, the
spin--parity--charge-conjugation assignment of such state reads
$J^{PC}=0^{-+}.$

The most general Salpeter amplitude $\phi(\bm{p})$ of a
\emph{spin-singlet\/} fermion--antifermion bound state is defined,
in its center-of-momentum frame, by two independent components
$\varphi_{1,2}(\bm{p})$:\begin{equation}
\phi(\bm{p})=\frac{1}{\sqrt{N_{\rm c}}}\left[\varphi_1(\bm{p})\,
\frac{H(\bm{p})}{E(\bm{p})}+\varphi_2(\bm{p})\right]\gamma_5\
.\label{Eq:PSA}\end{equation}In terms of these Salpeter
components, our bound-state normalization $N(P)$ introduced~in
Eq.~(\ref{Eq:BN}) becomes, in accordance with, {\it e.g.},
Refs.~\cite[Eq.~(4.13)]{L92I} and \cite[Eqs.~(18)
and~(26)]{OVW95},\begin{align}N(P)=(2\pi)^3\,\|\phi\|^2&
=4\int{\rm d}^3p\,[\varphi_1^\ast(\bm{p})\,\varphi_2(\bm{p})
+\varphi_2^\ast(\bm{p})\,\varphi_1(\bm{p})]\nonumber\\&
=4\int\limits_0^\infty{\rm d}p\,p^2\,
[\varphi_1^\ast(p)\,\varphi_2(p)+\varphi_2^\ast(p)\,\varphi_1(p)]\
,\qquad p\equiv|\bm{p}|\ ;\label{Eq:N}\end{align}here, the last
equality gives the norm in terms of the radial factors
$\varphi_{1,2}(p)$ obtained as relics of the Salpeter components
$\varphi_{1,2}(\bm{p})$ by factorizing out any dependence on
angular~variables.

In the interaction term (\ref{Eq:I}), the integral kernel
$K(\bm{p},\bm{q})$ acting on the Salpeter amplitude $\phi(\bm{p})$
reflects Lorentz nature and strength of the effective interactions
between bound-state constituents by generalized Dirac matrices
$\Gamma$ and Lorentz-scalar functions $V(\bm{p},\bm{q}).$ As~long
as not knowing better, we assume for fermion and antifermion
identical effective couplings, $$K(\bm{p},\bm{q})\,\phi(\bm{q})=
\sum_\Gamma V_\Gamma(\bm{p},\bm{q})\,\Gamma\,\phi(\bm{q})\,\Gamma\
,$$and the integral kernel to exhibit Fierz symmetry, realized,
for instance, by choosing for~the tensor structure
$\Gamma\otimes\Gamma$ a single linear combination of vector,
pseudoscalar, and~scalar terms:\begin{equation}\Gamma\otimes\Gamma
=\frac{1}{2}\,(\gamma_\mu\otimes\gamma^\mu+\gamma_5\otimes\gamma_5
-1\otimes1)\ .\label{Eq:D}\end{equation}Given that the interaction
term (\ref{Eq:I}) is of convolution type and respects spherical
symmetry, which is assured if
$K(\bm{p},\bm{q})=K((\bm{p}-\bm{q})^2)$ or
$V_\Gamma(\bm{p},\bm{q})=V_\Gamma((\bm{p}-\bm{q})^2)$ holds, our
bound-state equation (\ref{Eq:F}) with Dirac structure
(\ref{Eq:D}) can be boiled down to an eigenvalue problem with the
bound-state masses $\widehat M$ as its eigenvalues, posed by a
coupled system of one integral~and one algebraic equation, which
provides the radial factors $\varphi_{1,2}(p)$ in the components
$\varphi_{1,2}(\bm{p})$ \cite{WL07}:
\begin{subequations}\begin{align}&2\,E(p)\,\varphi_2(p)+
2\,Z^2(p^2)\int\limits_0^\infty\frac{{\rm d}q\,q^2}{(2\pi)^2}\,
V(p,q)\,\varphi_2(q)=\widehat M\,\varphi_1(p)\
,\label{Eq:5esa}\\&2\,E(p)\,\varphi_1(p)=\widehat M\,\varphi_2(p)\
,\label{Eq:5esb}\end{align}\end{subequations}with the interaction
potential $V(r)$ sought entering via its double Fourier--Bessel
transform$$V(p,q)\equiv\frac{8\pi}{p\,q}\int\limits_0^\infty{\rm
d}r\sin(p\,r)\sin(q\,r)\,V(r)\ ,\qquad q\equiv|\bm{q}|\ .$$When
solving the coupled equations (\ref{Eq:5esa}) and (\ref{Eq:5esb}),
we shall make a distinction of two~cases:\begin{itemize}\item For
$\widehat M=0,$ Eq.~(\ref{Eq:5esb}) implies $\varphi_1(\bm{p})=0,$
whence the Salpeter amplitude (\ref{Eq:PSA}) reduces~to
$$\phi(\bm{p})=\frac{1}{\sqrt{N_{\rm c}}}\,\varphi_2(\bm{p})\,
\gamma_5\ ,$$with the radial behaviour of its nonvanishing
component $\varphi_2(\bm{p})$ governed by Eq.~(\ref{Eq:5esa}):
\begin{equation}E(p)\,\varphi_2(p)+Z^2(p^2)\int\limits_0^\infty
\frac{{\rm d}q\,q^2}{(2\pi)^2}\,V(p,q)\,\varphi_2(q)=0\
.\label{Eq:M0}\end{equation}\item For $\widehat M\ne0,$ insertion
of one of the two relations (\ref{Eq:5esa}) and (\ref{Eq:5esb})
into the other~entails an eigenvalue problem for eigenvalues
${\widehat M}^2,$ equivalently defined by either of the~relations
\begin{subequations}\begin{align}4\,E^2(p)\,\varphi_1(p)
+4\,Z^2(p^2)\int\limits_0^\infty\frac{{\rm d}q\,q^2}{(2\pi)^2}\,
V(p,q)\,E(q)\,\varphi_1(q)&={\widehat M}^2\,\varphi_1(p)\
,\label{Eq:M2_1}\\4\,E^2(p)\,\varphi_2(p)+4\,Z^2(p^2)\,E(p)
\int\limits_0^\infty\frac{{\rm d}q\,q^2}{(2\pi)^2}\,V(p,q)\,
\varphi_2(q)&={\widehat M}^2\,\varphi_2(p)\ .\label{Eq:M2_2}
\end{align}\end{subequations}\end{itemize}Expressed, on an equal
footing, in terms of radial factor $\varphi_1(p)$ or
$\varphi_2(p)$ and abbreviating~by$$K_1^2\equiv
\int\limits_0^\infty{\rm d}p\,p^2\,E(p)\,|\varphi_1(p)|^2\ ,\qquad
K_2^2\equiv\int\limits_0^\infty{\rm d}p\,p^2\,
\frac{|\varphi_2(p)|^2}{E(p)}$$the arising integrals, the use of
relation (\ref{Eq:5esb}) translates the normalization condition
(\ref{Eq:N})~into$$16\,\frac{\operatorname{Re}(\widehat M)}
{|\widehat M|^2}\,K_1^2=4\,\operatorname{Re}(\widehat M)\,K_2^2=
N(P) \ .$$Here, we take into account the possibility that, without
established self-adjointness, the~set of solutions to static
reductions of the form (\ref{Eq:CMF}) might include non-real mass
eigenvalues~$\widehat M.$

\section{Lowest-Lying Pseudoscalar Mesons: Basic Features}
\label{Sec:MP}The mass eigenvalues of all states in the physical
sector of solutions must be real. Hence, for our intended analysis
of ground-state pseudoscalar mesons we feel safe to assume
$\widehat M^\ast=\widehat M.$ Retaining the normalization $N(P)$
in Eq.~(\ref{Eq:BN}) unspecified, we refrain from sticking to a
fixed choice of normalization and introduce, for arbitrary $N(P),$
the two features of ground-state pseudoscalar mesons that the
\emph{generalized\/} Gell-Mann--Oakes--Renner relation
\cite{MRT}~entwines:\begin{enumerate}\item The decay constant
$f_B$ for a generic pseudoscalar meson $B$ regarded as bound state
of quark and antiquark with field operators $\psi_{1,2}(X)$
parametrizes the vacuum-to-bound state matrix element of the weak
axial-vector current $A_\mu(X)
\equiv{:\!\bar\psi_1(X)\,\gamma_\mu\,\gamma_5\,\psi_2(X)\!:},$
$$\langle0|A_\mu(X)|B(P)\rangle={\rm i}\,\frac{\sqrt{N(P)}}
{(2\pi)^{3/2}\,\sqrt{2\,P_0}}\,f_B\,P_\mu \exp(-{\rm i}\,P\,X)\
,$$and is thus inferred by projecting the Salpeter amplitude of
$B$ onto the current~$A_0(0)$:$$f_B=\frac{{\rm
i}}{(2\pi)^{3/2}\,\sqrt{N(P)}}\,\sqrt{\frac{2}{\widehat
M}}\int{\rm d}^3p\,{\rm Tr}[\gamma_0\,\gamma_5\,\phi(\bm{p})]\ .$$
Use of Eq.~(\ref{Eq:PSA}) allows us to reshape $f_B$ to the
(trivially equivalent) more explicit~forms$$f_B=-\frac{{\rm
i}\,\sqrt{N_{\rm c}}}{\pi\,K_1}\int\limits_0^\infty{\rm
d}p\,p^2\,\frac{M(p^2)}{E(p)}\,\varphi_1(p)=-\frac{{\rm
i}\,\sqrt{N_{\rm c}}}{\pi\,K_2}\int\limits_0^\infty{\rm
d}p\,p^2\,\frac{M(p^2)}{E^2(p)}\,\varphi_2(p)\ .$$Both forms
clearly reveal that, for the specific Bethe--Salpeter model
considered~here, there is no explicit dependence of this decay
constant $f_B$ on the bound-state mass~$\widehat M.$\item
Likewise, we define, for a pseudoscalar meson $B,$ the associated
in-hadron condensate \cite{MRT} ${\mathbb C}_B$ by the
vacuum-to-bound state matrix element of the density
$:\!\bar\psi_1(0)\,\gamma_5\,\psi_2(0)\!:$,$$\frac{\sqrt{N(P)}}
{(2\pi)^{3/2}\,\sqrt{2\,P_0}}\,{\mathbb C}_B\equiv
\langle0|{:\!\bar\psi_1(0)\,\gamma_5\,\psi_2(0)\!:}|B(P)\rangle\
.$$Paralleling the $f_B$ case, this condensate is found by
projection onto the latter density,$${\mathbb
C}_B=-\sqrt{\frac{2\,\widehat M}{(2\pi)^3\,N(P)}}\int{\rm
d}^3p\,{\rm Tr}[\gamma_5\,\phi(\bm{p})]\ ,$$and, recalling
$\phi(\bm{p})$ of Eq.~(\ref{Eq:PSA}), reduces to the (clearly
equivalent) explicit expressions\footnote{For rather recent
determinations of in-hadron condensates rooted in the full
Bethe--Salpeter formalism augmented by the relevant
Dyson--Schwinger results, consult, for instance, Ref.~\cite{HGKL}
and references therein.}$${\mathbb C}_B=-\frac{2\,\sqrt{N_{\rm
c}}}{\pi\,K_1} \int\limits_0^\infty{\rm
d}p\,p^2\,E(p)\,\varphi_1(p) =-\frac{2\,\sqrt{N_{\rm
c}}}{\pi\,K_2}\int\limits_0^\infty{\rm d}p\,p^2\,\varphi_2(p)\
.$$\end{enumerate}With these definitions of decay constant $f_B$
and in-hadron condensate ${\mathbb C}_B$ of a pseudoscalar meson
$B$ and $m$ denoting the (by assumption, equal) mass parameters of
its constituents in the QCD Lagrangian, the modernized
Gell-Mann--Oakes--Renner relation of Ref.~\cite{MRT}~reads
\begin{equation}f_B\,\widehat M_B^2=2\,m\,{\mathbb C}_B\
,\label{Eq.:GOR}\end{equation}as a direct consequence of the
renormalized axial-vector Ward--Takahashi identity of QCD. (For
practical reasons, we suppress here any reference to issues
related to renormalization.)

\section{Inversion of Dyson--Schwinger Propagator Solution}
\label{Sec:I}At this stage, merely a sole ingredient to our
three-dimensional bound-state equation (\ref{Eq:CMF}) is still
lacking: the form of the Lorentz-scalar potential function
$V(\bm{p},\bm{q})$ capable of describing Goldstonic pseudoscalar
quark--antiquark bound states. As demonstrated in
Refs.~\cite{WL13,WL15,WL16:ARP,WL16:DSE,WL16i}, our bound-state
framework is sufficiently simple to allow, by means of inversion
techniques, for easy determination of these effective interactions
from some knowledge about solutions.

\begin{figure}[htp]\begin{center}\begin{tabular}{cc}
\psfig{figure=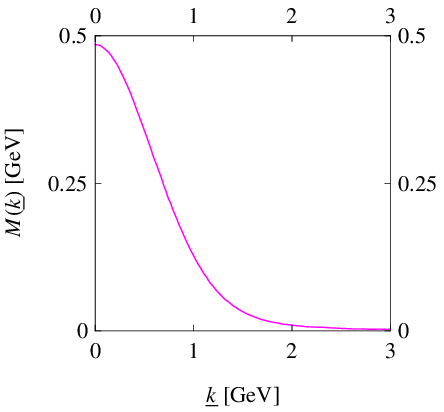,scale=1.748497}&
\psfig{figure=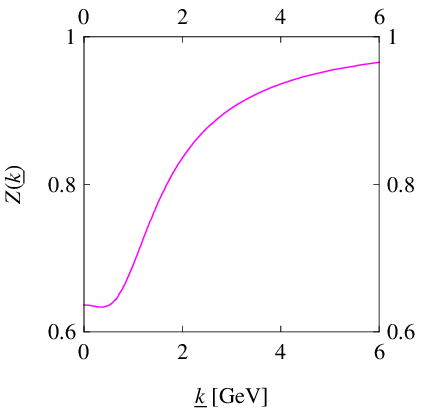,scale=1.748497}\\[1ex](a)&(b)\end{tabular}
\caption{Dyson--Schwinger solution \cite{PM99} for the full quark
propagator (\ref{Eq:P}) in the chiral~limit (depicted vs.\
$\underline{k}\equiv(\underline{k}^2)^{1/2}$ \cite{PM00}): (a)
mass $M(\underline{k})$ and (b) wave-function renormalization
$Z(\underline{k})$.}\label{Fig:MZI}\end{center}\end{figure}

\begin{figure}[htp]\begin{center}
\psfig{figure=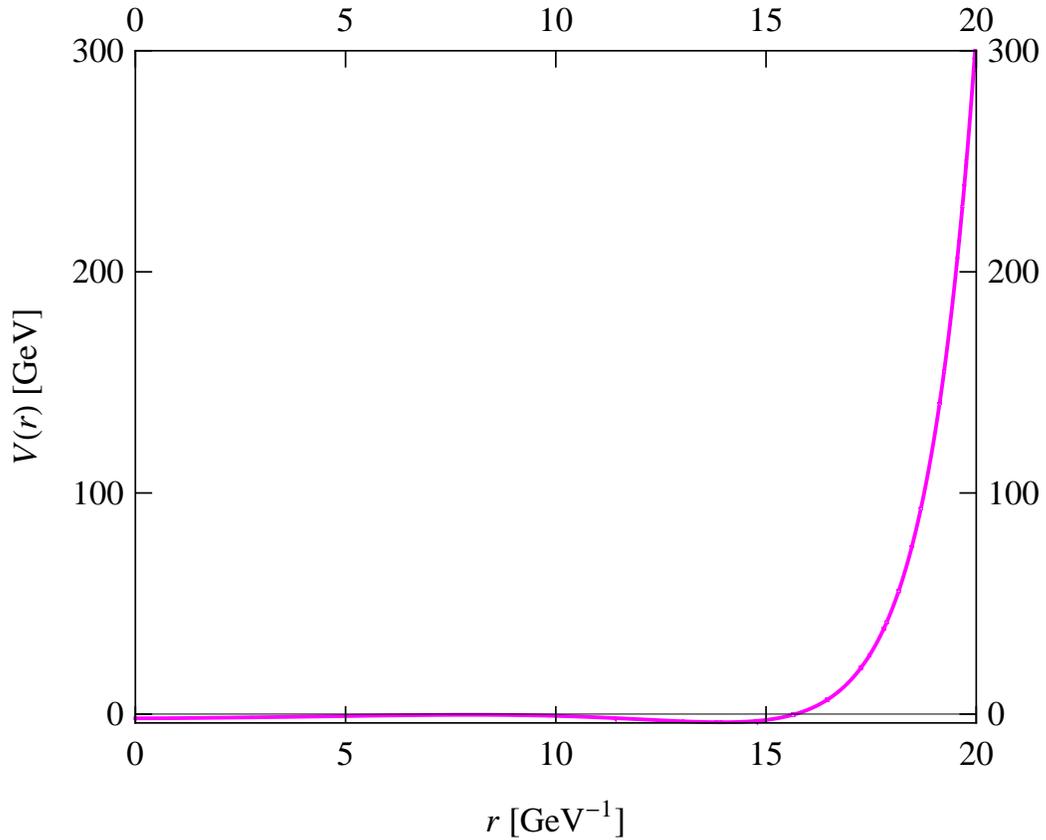,scale=2.0038}
\caption{Configuration-space potential $V(r)$ \cite{WL16i} found
upon inverting the Bethe--Salpeter problem (\ref{Eq:M0}) for a
Salpeter amplitude (\ref{Eq:PSA}) fixed by the quark propagator
functions of Fig.~\ref{Fig:MZI}.}\label{Fig:V}\end{center}
\end{figure}

In the chiral limit, definable, for QCD, by the vanishing of the
quark-mass parameter in its Lagrangian, the renormalized
axial-vector Ward--Takahashi identity of QCD relates the
center-of-momentum Bethe--Salpeter amplitude $\Phi(p,0)$ for the
massless flavour-nonsinglet pseudoscalar mesons to the quark
propagator $S(p)$ \cite{MRT}. In Euclidean-space representation,
indicated by underlining the corresponding variables, this
relation between Bethe--Salpeter amplitude and the quark
propagator functions $M(\underline{k}^2)$ and $Z(\underline{k}^2)$
can be reformulated as \cite{WL15}$$\Phi(\underline{k},0)\propto
\frac{Z(\underline{k}^2)\,M(\underline{k}^2)}
{\underline{k}^2+M^2(\underline{k}^2)}\,\underline{\gamma}_5+
\mbox{subleading contributions}\ .$$This relation can be
capitalized to convert, by inversion \cite{WL13}, knowledge about
the behaviour of the fermion propagator into information about the
underlying effective quark--antiquark interactions
\cite{WL13p,WL15,WL16:ARP,WL16:DSE,WL16i,WL16@,WL16c}. In
Ref.~\cite{WL16i}, exploiting the chiral-limit solution to the
Dyson--Schwinger equation for the quark propagator based on a
phenomenologically acceptable model for the four-point Green
function serving as required Bethe--Salpeter interaction-kernel
input \cite{PM99}, we extracted, from the quark-propagator
functions reproduced in Fig.~\ref{Fig:MZI}, the potential~$V(r)$
given in Fig.~\ref{Fig:V}, showing quark confinement by rising
from slightly negative $V(0)$~to infinity.

\section{Gell-Mann--Oakes--Renner-Related Characteristics}
\label{Sec:GOR}By use of, {\it e.g.}, the standard solution
methods sketched in App.~\ref{App}, it is now straightforward to
harvest the findings of our bound-state formalism for the
pseudoscalar-meson properties of interest. In order to track the
latter's behaviour with increasing quark mass $m,$ we mimic finite
values of $m$ by exploiting the propagator solutions provided in
Ref.~\cite[Fig.~1]{PM00} for the light quarks $q=u,d,s.$ The
predicted properties of the generic pseudoscalar meson~defined
thereby, collected in Table~\ref{Tab:fC}, exhibit satisfactory
agreement with the qualitative behaviour expected from the
Gell-Mann--Oakes--Renner-type relation (\ref{Eq.:GOR}): Within the
errors induced by the details of our treatment of the information
extracted pointwise from Ref.~\cite[Fig.~1]{PM00}, our
fictitious-meson mass squared $\widehat M_B^2$ vanishes in the
chiral limit $m=0$ and rises\footnote{The nature of this rise
cannot be determined from three data points for quark types
chiral, $u/d$ and~$s.$} with~$m.$ Moreover, we get a reasonable
proximity of the size of quark masses deduced from
Eq.~(\ref{Eq.:GOR}),$$m=\frac{f_B\,\widehat M_B^2}{2\,{\mathbb
C}_B}\ ,$$to the PDG averages \cite{PDG} (at a renormalization
scale $\mu=2\;\mbox{GeV}$) of the light current-quark masses
$\overline{m}_q(\mu)$ obtained in the modified minimal-subtraction
($\overline{\rm MS}$) renormalization~scheme,
$$\frac{\overline{m}_u+\overline{m}_d}{2}(2\;\mbox{GeV})
=\left(3.5^{+0.7}_{-0.3}\right)\mbox{MeV}\ ,\qquad
\overline{m}_s(2\;\mbox{GeV})=\left(96^{+8}_{-4}\right)\mbox{MeV}\
.$$

In summary, we conclude to have achieved the envisaged proof of
feasibility. Describing the lowest pseudoscalar mesons by an
advanced instantaneous Bethe--Salpeter equation \cite{WL05:LS}
with an effective interaction designed to reproduce these mesons'
Goldstone nature \cite{WL16i},~we gain numerical predictions for a
couple of fundamental properties of these quark--antiquark bound
states which, to say the least, are of the (experimentally)
correct order of magnitude and comply with the nature of their
interrelationship dictated by QCD on general~grounds.

\appendix\section{Matrix Representations of Bound-State Equations}
\label{App}The solutions to an explicit eigenvalue problem of the
kind posed by our radial bound-state equations (\ref{Eq:M2_1}) and
(\ref{Eq:M2_2}) can be, in principle, straightforwardly determined
by conversion to equivalent matrix eigenvalue problems,
accomplished by expansion of the eigenfunctions sought and, if
necessary, related quantities over some basis of the respective
function space. For instance, by expanding, in terms of basis
functions $\chi_i(r)$ in configuration space or
$\widetilde\chi_i(p)$ in momentum space ($i\in\mathbb{N}$), our
Salpeter function $\varphi_2(p)$, with coefficients $c_i$, and the
terms $Z^2(p^2)\,E(p)\,\widetilde\chi_i^*(p)$, the bound-state
equation (\ref{Eq:M2_2}) governing $\varphi_2(p)$ becomes the
eigenvalue equation of a matrix ${\mathbb O}=({\mathbb O}_{ij}),$
defined by kinetic elements ${\mathbb T}_{ij}$ and
potential~elements~${\mathbb V}_{kj}$:\begin{align*}
&\varphi_2(p)=\sum_{i=0}^{N<\infty}c_i\,\widetilde\chi_i(p)
\qquad\Longrightarrow\qquad\sum_{j=0}^{N<\infty}{\mathbb
O}_{ij}\,c_j={\widehat M}^2\,c_i\ ,\qquad{\mathbb
O}_{ij}=4\,{\mathbb T}_{ij}+4\sum_{k=0}^{N<\infty}d_{ik}\,{\mathbb
V}_{kj}\ ,\\&\quad{\mathbb T}_{ij}\equiv\int\limits_0^\infty{\rm
d}p\,p^2\,\widetilde\chi_i^*(p)\,E^2(p)\,\widetilde\chi_j(p)\
,\qquad{\mathbb V}_{ij}\equiv\int\limits_0^\infty{\rm
d}r\,r^2\,\chi_i^*(r)\,V(r)\,\chi_j(r)\ ,\\&\quad
d_{ij}\equiv\int\limits_0^\infty{\rm d}p\,p^2\,
\widetilde\chi_i^*(p)\,Z^2(p^2)\,E(p)\,\widetilde\chi_j(p)\qquad
\Longleftrightarrow\qquad Z^2(p^2)\,E(p)\,\widetilde\chi_i^*(p)
=\sum_{j=0}^{N<\infty}d_{ij}\,\widetilde\chi_j^*(p)\ .\end{align*}
Accordingly, it proves advantageous to employ a basis that may be
represented analytically in configuration and momentum space. In
the past \cite{WL:LB,WL:T1,WL:UA,WL:NV,WL:TWR,WL:WS,WL:H,WL:Y}, we
found it rather convenient to span the Hilbert space
$L_2(\mathbb{R}^+)$ of with weight $x^2$ square-integrable
functions on the positive real line $\mathbb{R}^+$ by an
orthonormalized basis that involves the generalized-Laguerre
orthogonal polynomials $L_i^{(\gamma)}(x)$ for parameter
$\gamma>-1$ \cite{AS,B}, and a variational
parameter~$\mu\in(0,\infty)$:
\begin{align*}\chi_i(r)&=\sqrt{\frac{(2\,\mu)^3\,i!}{\Gamma(i+3)}}
\exp(-\mu\,r)\,L_i^{(2)}(2\,\mu\,r)\ ,\qquad L_i^{(\gamma)}(x)
\equiv\sum_{t=0}^i\binom{i+\gamma}{i-t}\frac{(-x)^t}{t!}\ ,\\[1ex]
\widetilde\chi_i(p)&=4\,\sqrt{\frac{\mu^3\,i!}{\pi\,\Gamma(i+3)}}\,
\sum_{t=0}^i\,\frac{(-1)^t}{t!}\binom{i+2}{i-t}
\frac{\Gamma(t+2)\,(2\,\mu)^t}{p\,(p^2+\mu^2)^{t/2+1}}
\sin\!\left[(t+2)\arctan\frac{p}{\mu}\right],\\[1ex]
&\int\limits_0^\infty{\rm d}r\,r^2\,\chi_i(r)\,\chi_j(r)
=\int\limits_0^\infty{\rm d}p\,p^2\,\widetilde\chi_i^*(p)\,
\widetilde\chi_j(p)=\delta_{ij}\ ,\qquad i,j=0,1,2,\dots\
.\end{align*}

\begin{table}[t]\begin{center}\caption{Results for bound-state mass
$\widehat M,$ decay constant $f_B,$ and in-hadron condensate
${\mathbb C}_B$ of our description of lightest pseudoscalar mesons
defined by Eqs.~(\ref{Eq:F}) and (\ref{Eq:D}) together~with the
effective potential of Fig.~\ref{Fig:V}, for the three light-quark
scenarios of Ref.~\cite{PM99}, as well as~the quark mass parameter
$m$ fitting to the generalized Gell-Mann--Oakes--Renner relation
(\ref{Eq.:GOR}).}\label{Tab:fC}\vspace{2ex}
\begin{tabular}{rrccd{1.8}}\toprule\multicolumn{1}{c}{Constituents}&
\multicolumn{1}{c}{$\widehat M\;[\mbox{MeV}]$}&$f_B\;[\mbox{MeV}]$
&${\mathbb C}_B\;[\mbox{GeV}^2]$&
\multicolumn{1}{l}{$m\;[\mbox{MeV}]$}\\\midrule chiral quarks&
6.8&151&0.585&0.0059\\$u$/$d$ quarks&148.6&155&0.598&2.85\\$s$
quarks&
620.7&211&0.799&51.0\\\bottomrule\end{tabular}\end{center}\end{table}

\end{document}